\documentclass[a4paper,12pt,one column]{article}
\usepackage{graphicx}
\usepackage{colordvi}
\usepackage{amssymb}
\usepackage{latexsym}
\usepackage{hyperref}
\begin{document}
\title{Vacuum energy and Primordial Black Holes in Brans-Dicke Theory}
\author{D. Dwivedee$^{*}$, B. Nayak$^{\dag}$ and L. P. Singh$^{\ddag}$ \\
Department of Physics, Utkal University,
Bhubaneswar 751004, India. \\
$^{*}$debabrata@iopb.res.in \\ $^{\dag}$bibeka@iopb.res.in \\ $^{\ddag}$lambodar$\_$uu@yahoo.co.in \\}
\date{}
\maketitle
\begin{abstract}
In our work, we study the evolution of primordial black holes within the context of Brans-Dicke theory by assuming present universe as vacuum dominated. We also consider the accretion of radiation, matter and vacuum energy only during respective dominant periods. From our study, we found that the accretion rate is slower in both radiation and vacuum energy dominated eras in Brans-Dicke theory in comparision with General Theory of Relativity \cite{nj}. Thus the PBHs evaporate at a faster rate in Brans-Dicke theory than Standard Cosmology \cite{nj}, if we consider the presence of vacuum energy in both cases. We also find that vacuum energy accretion efficiency should be less than $0.61$. 
\end{abstract}

\section{Introduction}
In reference \cite{nj}, by taking General Theory of Relativity (GTR) \cite{ein} as the theory of gravity, it is shown that during vacuum dominated era, the accretion of vacuum energy increases the mass of primordial black holes. Here we try to extend this work by changing the theory of gravity from GTR to scalar-tensor theory like Brans-Dicke (BD) theory \cite{bdt}. Due to the time variation of Newton's gravitational constant $G$ in Brans-Dicke theory, the scale factor takes a different form and it controls the PBH evolution in a different manner compared with GTR. So it is worthwhile to study PBH evolution in vacuum dominated era within Brans-Dicke theory.

Einstein's General Theory of Relativity is based on a pure tensor
theory of gravity where gravitational constant is taken as a
time-independent quantity. But Brans-Dicke theory is a scalar-tensor
theory of gravity where the gravitational constant is a time-dependent
quantity. BD theory is the simplest extension over GTR through the
introduction of a time-dependent scalar field $\phi(t)$ as $G(t)$ $\sim $
$\phi^{-1}(t)$, where the scalar field $\phi(t)$ couples to gravity with a
coupling parameter $\omega$ known as the BD parameter. BD theory goes over
 to GTR in the limit $\omega$ $\rightarrow$
$\infty$ \cite{bam1, bam2}. BD type model can also be regarded as the low energy limit of
Kaluza-Klein and String theories \cite{asm1, asm2, asm3}. Again BD theory explains many cosmological
phenomena such as inflation \cite{johri, la}, early and late time behaviour of the universe \cite{sahoo1, sahoo2}, cosmic acceleration and structure formation \cite{bermar}, coincidence
problem \cite{nayak} and problems relating to black holes \cite{ijtp}.

                           Primordial Black Holes (PBHs) are those black
holes which are formed in the early universe through variety of mechanisms
such as inflation \cite{cgl, kmz}, initial inhomogeneities \cite{carr, swh}, phase transition and critical phenomena in gravitational collapse \cite{khopol1, khopol2, khopol3, khopol4, jedam1, jedam2, jedam3}, bubble collision \cite{kss} or the
decay of cosmic loops \cite{polzem1, polzem2}. A comparision of cosmological density of the
universe with the density associated with a black hole, at any time after
BigBang, shows that formation mass of PBH would have same order as that of horizon mass. Thus PBH could span wide mass range starting from Planck mass
$10^{-5} g$ to more than $ 10^{15} g$. Hawking has also shown that black
holes can emit thermal radiation quantum mechanically \cite{hawk}. So black holes will
evaporate depending upon their formation masses. Smaller the mass of
PBHs quicker they evaporate. As density of a black hole varies inversely
with its mass, high density which is possible in the early universe, is
required to form lighter black holes. So PBHs  with very small mass in comparision with their stellar or galactic counterparts can evaporate completely by the present epoch through Hawking evaporation \cite{hawk}. Early evaporating PBHs could account for baryogenesis \cite{bckl, mds1, mds2, dol1, dol2}. On the
other hand, Longer lived PBHs could act as seeds for structure formation \cite{mor1, mor2, mor3, mor4, mor5, mor6}
and could also form a significant component of the dark matter \cite{blais1, blais2, blais3, blais4}. We have also shown that \cite{nsmprd1, nsmprd2, pramana},
PBHs could take comparatively more time to evaporate due to accretion of
radiation which makes them long lived.

The standard picture of cosmology states that the universe is
radiation dominated in the very beginning of its evolution and becomes
matter dominated at a later stage. The expansion of the universe is decelerated throughout its evolution in both the periods. But the observations of distant Supornovae of type Ia (SNIa) \cite{apj1, apj2, apj3} indicate that the expansion of universe is accelerating in the present epoch. This has led to the conclusion that nearly two-third of the critical energy density of the universe exists in a dark energy component with a large
negative pressure and of unknown composition. The simplest candidate for the dark energy is vacuum energy with equation of state parameter $\gamma=-1$. Recent Planck data shows that dark energy
occupies $68.3\%$ of universe and the rest $31.7\%$ is contributed by dark and luminous form of matter. SNIa observations also provide the evidence of transition from decelerated to accelerated phase occuring at redshift $z_{q=0}\sim0.46$ \cite{mst1, mst2}. So the vacuum energy should dominate from $z_{q=0}\sim 0.46$. 

In the present study, we integrate vacuum energy accretion by PBHs with radiation and matter accretions in respective dominant periods and its effect on their evolution. We also present  comparision of the results of the present study with the corresponding results of the previous work using  Standard Cosmology \cite{nj}.

\section{Primordial Black Holes and Brans-Dicke Theory}

For a spatially flat FRW universe with scale factor $a$, the Friedmann equations and the equation of motion for BD field $\phi$ are given by
\begin{eqnarray}
\frac{{\dot{a}}^{2}}{a^{2}}+\frac{\dot{a}}{a}\frac{\dot{\phi}}{\phi}
-\frac{\omega}{6}\frac{{\dot{\phi}}^{2}}{{\phi}^{2}}=\frac{8\pi
\rho}{3\phi}
\end{eqnarray}
\begin{eqnarray}
2\frac{\ddot{a}}{a}+\frac{\dot{a}^2}{a^2}+2\frac{\dot{a}}{a}\frac{\dot{\phi}}{\phi}+\frac{\omega}{2}\frac{\dot{\phi}^2}{\phi^2}+\frac{\ddot{\phi}}{\phi}=-\frac{8\pi p}{\phi}
\end{eqnarray}
\begin{eqnarray}
\frac{\ddot{\phi}}{8\pi}+3\frac{\dot{a}}{a}\frac{\dot{\phi}}{8\pi}=\frac{\rho-3p}{2\omega+3}
\end{eqnarray}
where $\rho$ and $p$ denote the total energy density and pressure of the fluid filling the universe.

The energy conservation equation can now be written as
\begin{equation}
\dot{\rho}+3(\gamma+1)H\rho=0
\end{equation}
where $H=\frac{\dot a}{a}$ is the Hubble parameter and $\gamma$ is the equation of state parameter taking values 1/3 for radiation, 0 for matter and -1 for vacuum energy. The universe evolves through radiation $(t<t_1)$, matter $(t_1<t<t_2)$ and vacuum dominated era $(t>t_2)$.

From equation (4), we find \cite{swg, swc}
\begin{eqnarray}
\rho(a)\propto \left\{
\begin{array}{rrr}
a^{-4} & (t<t_1)\\
a^{-3} & (t_1<t<t_2)\\
c_1 & (t>t_2)
\end{array}
\right.
\label{7}
\end{eqnarray}

where $t_1$ is the time of matter-radiation equality and $t_2$ is the time after which vacuum energy dominates.
We use the solutions of $G(t)$ and $a(t)$ for radiation, matter and vacuum dominated eras  obtained by Barrow and Carr \cite{jdbj} and by us \cite{dbl} as
\begin{eqnarray}
G(t)= \left\{
\begin{array}{rrr}
G_0\Big(\frac{t_0}{t_1}\Big)^{n} & (t<t_1)\\
G_0\Big(\frac{t_0}{t}\Big)^{n} & (t_1<t<t_2)\\
G_0\Big(\frac{t_0}{t}\Big)^{2} & (t>t_2)
\end{array}
\right.
\label{7}
\end{eqnarray}

and

\begin{eqnarray}
a(t)\propto \left\{
\begin{array}{rr}
t^{\frac{1}{2}} & (t<t_1)\\
t^{\frac{2-n}{3}} & (t_1<t<t_2)\\
t^{\Big(-1+\sqrt{1+\frac{8\pi G_0}{3} t_0^2 \rho_{v} +\frac{2}{3}\omega}\Big)} & (t>t_2)
\end{array}
\right.
\label{7}
\end{eqnarray}

where $G_0$ is the present value of Newton's gravitational constant $G$, $t_0$ is the present time and $n$ is a parameter related to $\omega$ as $n=\frac{2}{4+3\omega}$. Again solar system observations \cite{blp} require that $\omega$ be large $(\omega \geq 10^{4})$ and hence $n$ is very small $(n\leq 0.00007)$.



Due to Hawking radiation, PBH mass decreases at a rate given by
\begin{equation}
\dot{M}(t)_{evap}=-\frac{a_H}{256\pi^3}\frac{1}{G^2M^2_{evap}} 
\end{equation}
where $a_H$ is the Stefan-Boltzmann constant multiplied with number of degrees of freedom available for radiation and $M_{evap}$ represents evolution of PBH mass due to Hawking radiation process only.

PBH mass, however,  can  change due to the accretion of radiation, matter or vacuum energy at a rate given by
\begin{equation}
\dot {M}(t)_{acc}=16\pi G^2f_jM^2_{acc}\rho_j  
\end{equation}
where $M_{acc}$ denotes evolution of PBH mass due to accretion only, $f_j$ and $\rho_j$ denote the accretion efficiency and density respectively of the dominant energies, denoted by $j$, in different eras. The value of accretion efficiency $f_j$ depends upon the complex physical processes such as the mean free paths of the particles comprising the surroundings of the PBHs. Any peculiar velocity of the PBH with respect to the cosmic frame could increase the value of $f_j$ \cite{apr, npr}.

In our calculation, we use the numerical values for different quantities like $G_0=6.67\times 10^{-8}dyn.cm^2/g^2$, $\rho_{cr}=1.1\times 10^{-29} g/cm^{3}$, $t_1=10^{11}s$, and vacuum energy density $\rho_v=0.683\times \rho_{cr}$. Now we can calculate the numerical value of $t_2$ by using the data $z_{q=0}=0.46$ \cite{mst1, mst2}.\\ 
From the definition of red shift, we have
\begin{equation}
1+z_{q=0}=\frac{a(t_0)}{a(t_2)} 
\end{equation}
Using equation (7) and value of $z_{q=0}$, we get
\begin{equation}
\Big(\frac{t_0}{t_2}\Big)^{(-1+\sqrt{1+\frac{8\pi}{3}G_0t_0^2\rho_v+\frac{2}{3}\omega})}=1.46 
\end{equation}
Taking the numerical values of different quantities we obtain $t_2=0.995\times t_0$ with $t_0=4.42\times 10^{17}s$.

\section{Study of Accretion regimes}

In this section, we study only accretion neglecting Hawking evaporation in order to clarify the effect of different accretions on the mass evolution of PBHs. 

\subsection{Accretion of radiation $(t<t_1)$}
When a PBH immersed in radiation field, the accretion of radiation leads to increase of its mass at a rate given by

\begin{equation}
\dot{M}(t)_{acc}=16\pi G^2 f_{rad}M^2_{acc}\rho_r
\end{equation}
Taking the solutions of $G(t)$ from equation (6) and $a(t)$ from equation (7), we can obtain from equation (1) that
\begin{equation}
\rho_r(t)=\frac{3}{32\pi G_0}\Big(\frac{t_1}{t_0}\Big)^n\frac{1}{t^2}
\end{equation}
Using equations (6) and (13) in equation (12), we get
\begin{equation}
\dot{M}(t)_{acc}=\frac{3}{2}f_{rad}G_0\Big(\frac{t_0}{t_1}\Big)^n\frac{M^2_{acc}}{t^2}
\end{equation}
On integration, equation (14) leads to
\begin{equation}
M(t)_{acc}=M_i\Big[1+\frac{3}{2}f_{rad}G_0\Big(\frac{t_0}{t_1}\Big)^nM_i\Big(\frac{1}{t}-\frac{1}{t_i}\Big)\Big]^{-1}
\end{equation}
where $M_i$ is the initial mass of PBH at time $t_i$ in radiation dominated era.

Assuming that PBHs would have mass of the order of the horizon mass at their formation epoch, we write $ M_i(t_i)= M_H(t_i)=\Big[G_0\Big(\frac{t_0}{t_1}\Big)^n\Big]^{-1}t_i$.
With this value of $M_i$, equation (15) becomes
\begin{equation}
M(t)_{acc}=M_i\Big[1+\frac{3}{2}f_{rad}\Big(\frac{t_i}{t}-1\Big)\Big]^{-1}
\end{equation}
For large time t, this equation reduces to
\begin{equation}
M(t)_{acc}=\frac{M_i}{1-\frac{3}{2}f_{rad}}
\end{equation}

For the $M(t)_{acc}$ to be positive, the radiation accretion efficiency needs to be bound like $f_{rad} <\frac{2}{3}$ in contrast to a value $f_{rad}<0.366$ of standard cosmology \cite{nj}.     
We have shown the variation of PBH mass with time only due to accretion in figure-1 for three different values of radiation accretion efficiency $f_{rad}$ as $0.05$, $0.1$ and $ 0.15$.

\begin{figure}[h]
\centering
\includegraphics[scale=0.8]{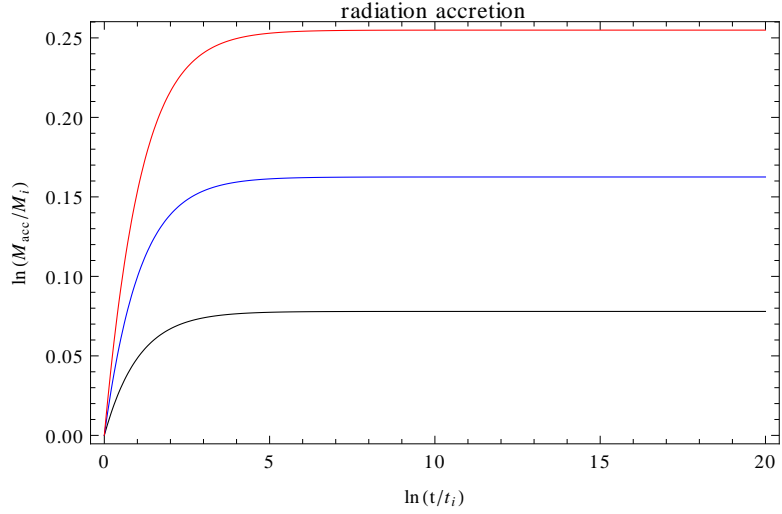}
\caption{Variation of PBH mass with time for different radiation accretion efficiencies $f_{rad}$ as $0.05$ (black), $0.1$ (blue) and $ 0.15$ (red).}
\end{figure} 

The figure-1 shows that the mass of the PBH increases with increase in radiation accretion efficiency and for a particular value of accretion efficiency mass of PBH saturates after a brief period of growth as in Standard Cosmology. But, here, in BD theory the radiation accretion rate is slower compared to GTR \cite{nj}.
\subsection{Accretion of matter $(t_1<t<t_2)$}
In matter dominated era, a PBH accretes surrounding matter for which its mass increases at a rate given by
\begin{equation}
\dot{M}(t)_{acc}=16\pi G^2 f_{mat}M^2_{acc}\rho_m
\end{equation}
where $\rho_m$ denotes matter density.\\
Using equations (6) and (7) in equation (1) with $\phi=G^{-1}(t)$, we get
\begin{equation}
\rho_m(t)=\frac{3}{8\pi G_0t_0^n}\Big[\frac{4}{9}+\frac{2}{9}n-\frac{4+3\omega}{18}n^2\Big]t^{n-2}
\end{equation}
Equation (18) can now be modified by the use of equations (6) and (19) as
\begin{eqnarray}
\dot{M}(t)_{acc}=\Big[\frac{8}{3}+\frac{4}{3}n-\frac{4+3\omega}{3}n^2\Big]G_0t_0^nf_{mat}\frac{M^2_{acc}}{t^{n+2}} 
\end{eqnarray}
The solution of equation (20) gives
\begin{eqnarray}
M(t)_{acc}=M_i\Big[1+\Big\{\frac{8}{3(n+1)}+\frac{4}{3}\frac{n}{n+1}-\frac{4+3{\omega}}{3}\frac{n^2}{n+1}\Big\}G_0t_0^nf_{mat}M_i(t^{-n-1}-t_i^{-n-1})\Big]^{-1}
\end{eqnarray}
Taking horizon mass as initial mass of PBH in matter dominated era i.e. $M_i(t_i)=[G_0(\frac{t_0}{t_i})^n]^{-1}t_i$, we can write

\begin{equation}
M(t)_{acc}=M_i\Big[1+\Big\{\frac{8}{3(n+1)}+\frac{4}{3}\frac{n}{n+1}-\frac{4+3\omega}{3}\frac{n^2}{n+1}\Big\}f_{mat}\Big\{\Big(\frac{t_i}{t}\Big)^{n+1}-1\Big\}\Big]^{-1}
\end{equation}
Since $n$ is small, for large time the above equation becomes
\begin{equation}
M(t)_{acc}=M_i[1-\frac{8}{3}f_{mat}]^{-1}
\end{equation}
The validity of equation (23) requires $f_{mat}<\frac{3}{8}$.
The variation of accreting mass with time for different matter accretion efficiencies is shown in figure-2.
\begin{figure}[h]
\centering
\includegraphics[scale=0.8]{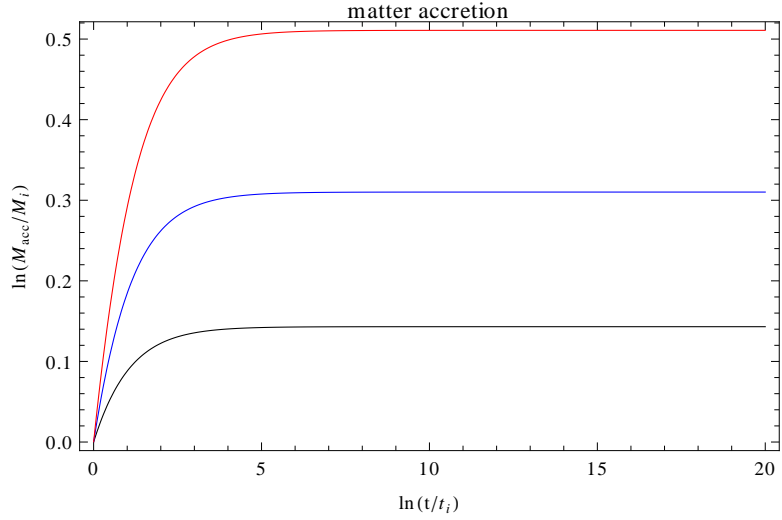}
\caption{Variation of PBH mass with time for different matter accretion efficiencies $f_{mat}$ as $0.05$ (black), $0.1$ (blue) and $ 0.15$ (red).}
\end{figure}

Figure-2 indicates that with increase in accretion efficiency PBH mass increases
due to matter accretion as Standard Cosmology \cite{nj}, but here accretion rate is slightly higher than that in Standard case. 
\subsection{Accretion of vacuum energy $(t>t_2)$}
In vacuum dominated era, the  mass of PBH is affected by vacuum energy at a rate given by
\begin{equation}
\dot{M}(t)_{acc}=16\pi G^2f_{vac}M^2_{acc}\rho_v
\end{equation}
where $\rho_v=\Omega^0_{\Lambda}\rho_{cr}$ with $\Omega^0_{\Lambda}=0.683$ is the present cosmological vacuum energy density parameter and $f_{vac}$ denotes vacuum energy accretion efficiency of PBH.

Now using equation (6) for $G(t)$ and the above expression for $\rho_v$, we can write equation (24) as
\begin{equation}
\dot{M}(t)_{acc}=16\pi G_0^2{t_0}^4 f_{vac} \Omega^0_{\Lambda}\rho_{cr} \frac{M^2_{acc}}{t^4}
\end{equation}
By integrating equation (25), we can obtain
\begin{equation}
M(t)_{acc}=M_i\Big[1+\frac{16\pi}{3}G_0^2t_0^4f_{vac}\Omega^0_{\Lambda}\rho_{cr}M_i\Big(\frac{1}{t^3}-\frac{1}{t_i^3}\Big)\Big]^{-1}
\end{equation}

Since the horizon mass varies with time as initial mass of PBH, so here $M_i(t_i)=\Big[G_0\Big(\frac{t_0}{t_i}\Big)^2\Big]^{-1}t_i$.
Using this value of $M_i$, one can find
\begin{equation}
M(t)_{acc}=M_i\Big[1+\frac{16\pi}{3}G_0t_0^2f_{vac}\Omega^0_{\Lambda}\rho_{cr}\Big\{\Big(\frac{t_i}{t}\Big)^3-1\Big\}\Big]^{-1}
\end{equation}
Substituting the numerical values of different quantities, we get

\begin{equation}
M(t)_{acc}=M_i\Big[1+1.639f_{vac}\Big\{\Big(\frac{t_i}{t}\Big)^3-1\Big\}\Big]^{-1}
\end{equation}
For large time, $\frac{t_i}{t}\to 0$ and we can write
\begin{equation}
M(t)_{acc}=M_i[1-1.639f_{vac}]^{-1}
\end{equation}
The validity of above equation gives $f_{vac}<\frac{1}{1.639} \approx0.61$.
The variation of PBH mass with time due to only vacuum energy accretion is shown in figure-3.

\begin{figure}[h]
\centering
\includegraphics[scale=0.8]{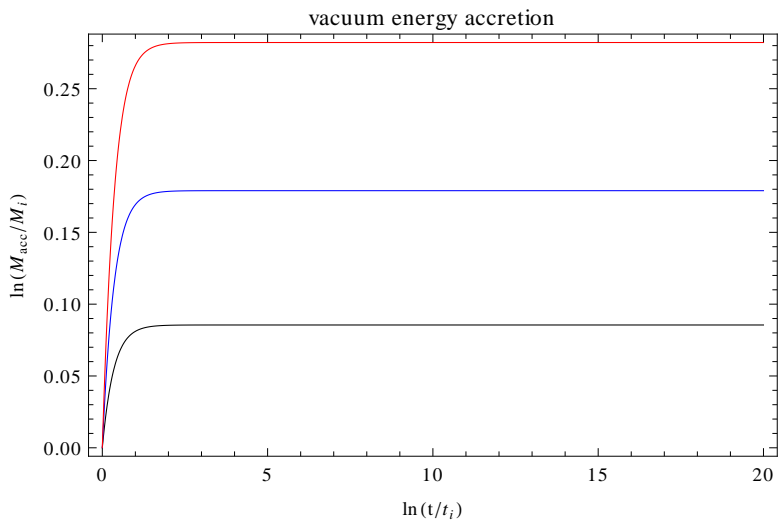}
\caption{Variation of PBH mass with time for different vacuum energy accretion efficiencies $f_{vac}$ as $0.05$ (black), $0.1$ (blue) and $ 0.15$ (red).}
\end{figure}

The comparision of Fig-3 with the corresponding figure of Standard Cosmology analysis \cite{nj} shows the interesting difference that where as in Standard Cosmology PBH mass saturates at a cut-off time and has no evolution beyond it, in the BD case there is a cut-off mass beyond which it never grows with time. Further, in BD case the rate of growth of PBH mass is slower compared with the standard case.

\section{Evolution of PBH in different eras}
We, now, study the evolution of PBH formed in radiation dominated era by taking both accretion and evaporation into account.
\subsection{Radiation dominated era}
Using equations (8) and (14), we write the variation of PBH mass in radiation dominated era as
\begin{equation}
\dot{M}(t)_{PBH}=\frac{3}{2}f_{rad}G_0\Big(\frac{t_0}{t_1}\Big)^n\frac{M^2_{PBH}}{t^2}-\frac{a_H}{256\pi^3}\Big(\frac{t_1}{t_0}\Big)^{2n}\frac{1}{G_0^2M_{PBH}^2}
\end{equation}
where $M_{PBH}$ represents a real evolution of PBH mass by considering both accretion and evaporation effects.
As the above equation (30) can not be solved analytically, we use numerical method to solve it. From the results we construct Table-1 to show the evaporation time of a particular PBH which is formed in radiation dominated era. 

\begin{table}[h]
\centering
\begin{tabular}[c]{|c|c|}
\hline
\multicolumn{2}{|c|}{${t_i =10^{-27}s}$ and $M_i=10^{11}g$}\\
\hline
$f_{rad}$ & $(t_{evap})$ \\
\hline
0 & $3.333\times10^{4}s$ \\
\hline
0.1 & $5.428\times10^{4}s$\\
\hline
0.2 & $9.718\times10^{4}s$ \\
\hline
0.3 & $2.004\times10^{5}s$\\
\hline
0.4 & $5.208\times10^{5}s$ \\
\hline
\end{tabular}
\caption{ Evaporation times of PBHs which are formed in radiation-dominated era  for different radiation accretion efficiencies.}
\end{table}

Table-1 shows that the increase of accretion efficiency prolongates the life time of PBHs in BD theory as in Standard Cosmology. But the BD PBHs evaporate at a faster rate than Standard case \cite{nj} due to the slower rate of accretion of radiation as was observed earlier while discussing Fig-1.

\subsection{Matter dominated era}
PBHs are usually not formed in matter dominated era. We, here, consider the PBH which is formed in radiation dominated era.

The equation governing the variation of PBH mass with time, in matter dominated era, can be written by using equations (8) and (20) as
\begin{equation}
\dot{M}(t)_{PBH}=\Big[\frac{8}{3}+\frac{4}{3}n-\Big(\frac{4+3\omega}{3}\Big)n^2\Big]G_0t_0^nf_{mat}\frac{M^2_{PBH}}{t^{n+2}}-\frac{a_H}{256\pi^3}\frac{1}{G_0^2t_0^{2n}}\frac{t^{2n}}{M^2_{PBH}}
\end{equation}

Solving equation (31) along with equation (30) numerically, we construct Table-2 indicating evaporation time of a particular PBH formed in radiation dominated era.

\begin{table}[h]
\centering
\begin{tabular}[c]{|c|c|}
\hline
\multicolumn{2}{|c|}{${t_i =10^{-25}s}$, ${M_i=10^{13}g}$ and $f_{rad}=0.35$}\\
\hline
$f_{mat}$ & $(t_{evap})$ \\
\hline
0 & $3.11\times10^{11}s$ \\
\hline
0.05 & $3.11\times10^{11}s$ \\
\hline
0.15 & $3.11\times10^{11}s$ \\
\hline
0.25 & $3.11\times10^{11}s$ \\
\hline
0.35 & $3.11\times10^{11}s$ \\
\hline
\end{tabular}
\caption{ Evaporation times of PBHs which are formed in radiation-dominated era  but evaporated in matter dominated era with different matter accretion efficiencies.}
\end{table}

Table-2 shows that the PBH evolution is not affected by accretion of matter. Similar kind of results was obtained in Standard Cosmology \cite{nj} though evaporation rate in BD theory is faster compared with the Standard case due to the slower rate of radiation accretion as found earlier. 
\subsection{Vacuum energy dominated era}
The rate of variation of PBH mass in vacuum dominated era is given by taking equations (8) and (25) as
\begin{equation}
\dot{M}(t)_{PBH}=16\pi G^2_0t^4_0f_{vac}\Omega^0_{\Lambda}\rho_{cr}\frac{M^2_{PBH}}{t^4}-\frac{a_H}{256\pi^3}\frac{1}{G^2_0t^4_0}\frac{t^4}{M^2_{PBH}}
\end{equation}

Solving numerically the above equation (32) along with equations (30) and (31), we construct Table-3 for a particular PBH evaporating in vacuum energy dominated era. 
\begin{table}[h]
\centering
\begin{tabular}[c]{|c|c|}
\hline
\multicolumn{2}{|c|}{${t_i =10^{10}s}$, $M_i=10^{48}g$, $f_{rad}=0.6$ and $f_{mat}=0.35$}\\
\hline
$f_{vac}$ & $(t_{evap})$ \\
\hline
0 & $9.3227\times10^{37}s$ \\
\hline
0.2 & $9.3227\times10^{37}s$ \\
\hline
0.4 & $9.3227\times10^{37}s$ \\
\hline
0.6 & $9.3227\times10^{37}s$ \\
\hline
\end{tabular}
\caption{ Evaporation times of PBHs which are formed in radiation-dominated era  but evaporated in vacuum dominated era for different values of vacuum energy accretion efficiency.}
\end{table}

It is found from Table-3 that accretion of vacuum energy has insignificant effect on the life span of PBH in contrast to Standrd Cosmology \cite{nj}. So PBHs evaporated at a faster rate in BD theory than Standard case \cite{nj}.

\section{Constraints on mass of PBH}
As observed astrophysical constraints arise from the presently evaporating PBHs, we, here, discuss about the PBHs whose evaporation time is $t_0$. We, now, calculate the initial mass of these PBHs, denoted by $(M_i)_{vac}$, in the presence of vacuum energy and also the initial mass of these PBHs, denoted by $M_i$, by not considering vacuum energy. For the calculation of $(M_i)_{vac}$ we use the numerical solutions of eqns (30), (31) and (32) and for $M_i$ we use numerical solutions of eqns (30) and (31). The results are shown in Table-4.

\begin{table}[h]
\centering
\begin{tabular}[c]{|c|c|c|}
\hline
\multicolumn{3}{|c|}{${t_{evap}=t_0=4.42\times10^{17}s}$ and $f_{mat}=0.35$}\\
\hline
$f_{rad}$ & $M_i$ & $(M_i)_{vac}$ \\
\hline
0 & $2.3669\times10^{15}g$ & $2.3669\times10^{15}g$\\
\hline
0.2 & $1.65685\times10^{15}g$ & $1.65685\times10^{15}g$\\
\hline
0.4 & $0.94678\times10^{15}g$ & $0.94678\times10^{15}g$\\
\hline
0.6 & $0.23669\times10^{15}g$ & $0.23669\times10^{15}g$\\
\hline
\end{tabular}
\caption{Formation masses of the PBHs which are evaporating now with different accretion efficiencies by considering both vacuum energy domination $(M_i)_{vac}$ and not vacuum energy domination $(M_i)$.}
\end{table}

Results of Table-4 indicate that the vacuum energy accretion does not affect the lifetimes of presently evaporating PBHs. The same is also true for PBHs which have completely evaporated by the present time.

Now we estimate the constraint arsing from the present $\gamma$-ray background \cite{jb, ysk, agl} to impose limits on the initial mass spectrum of PBHs in BD theory.

The fraction of the universe's mass going into PBHs at time t is given by \cite{bjca, jdbj}
\begin{equation}    
\beta (t)=\Big[\frac{\Omega_{PBH}(t)}{\Omega_R}\Big](1+z)^{-1}
\end{equation}
where $\Omega_{PBH}(t)$ represents the present density parameter associated with PBHs formed at time t, $z$ represents the redshift associated with time t and $\Omega_R=10^{-4}$ represents the present microwave background density. As the presently evaporated PBHs are formed in radiation dominated era $(t<t_1)$, one can obtain from the definition of red shift that
\begin{equation}
(1+z)^{-1}=\Big(\frac{t}{t_1}\Big)^{\frac{1}{2}} \Big(\frac{t_1}{t_2}\Big)^{\frac{2-n}{3}}\Big(\frac{t_2}{t_0}\Big)^{(-1+\sqrt{1+\frac{8\pi}{3}G_0t_0^2\rho_v+\frac{2}{3}\omega})}
\end{equation}
Now using equation (34) and the value of $\Omega_R$ we can write equation (33) as 
\begin{equation}
\beta (t)=\Big(\frac{t}{t_1}\Big)^{\frac{1}{2}} \Big(\frac{t_1}{t_2}\Big)^{\frac{2-n}{3}}\Big(\frac{t_2}{t_0}\Big)^{(-1+\sqrt{1+\frac{8\pi}{3}G_0t_0^2\rho_v+\frac{2}{3}\omega})}\Omega_{PBH}(t)\times 10^4
\end{equation}
Using horizon mass as the formation mass of PBH, one can find for radiation dominated era that $M=G^{-1}t=G_0^{-1}\Big(\frac{t_1}{t_0}\Big)^n t$.
Thus we can write the fraction of the universe mass going into PBHs as a function of the formation mass (M) as

\begin{equation}
\beta (M)=\Big(\frac{M}{M_1}\Big)^{\frac{1}{2}} \Big(\frac{t_1}{t_2}\Big)^{\frac{2-n}{3}}\Big(\frac{t_2}{t_0}\Big)^{(-1+\sqrt{1+\frac{8\pi}{3}G_0t_0^2\rho_v+\frac{2}{3}\omega})}\Omega_{PBH}(M)\times 10^4
\end{equation}
where $M_1=M(t_1)$.

Observations imply that $\Omega_{PBH}(M)<1$ over all mass ranges of PBHs living beyond the present time. But presently evaporating PBHs generate a $\gamma$-ray background with most of the energy around $100 MeV$ \cite{idn}. If $\epsilon_{\gamma}$ is the fraction of the emitted energy going into photons then the density of the radiation at this energy is expected to be $\Omega_{\gamma}=\epsilon_{\gamma}\Omega_{PBH}(M_*)$; where $M_*$ represents the formation mass of presently evaporating PBHs. since $\epsilon_{\gamma}=0.1$ \cite{epsl} and the observed $\gamma$-ray background density around $100 MeV$ is $\Omega_{\gamma} \sim 10^{-9}$ \cite{jb, ysk, agl}, we get $\Omega_{PBH}(M_*)<10^{-8}$ .

Using this limit of $\Omega_{PBH}$ for $M=M_*$, equationn (36) leads to an upper bound 
\begin{equation}
\beta (M_*)<\Big(\frac{M_*}{M_1}\Big)^{\frac{1}{2}} \Big(\frac{t_1}{t_2}\Big)^{\frac{2-n}{3}}\Big(\frac{t_2}{t_0}\Big)^{(-1+\sqrt{1+\frac{8\pi}{3}G_0t_0^2\rho_v+\frac{2}{3}\omega})}\times 10^{-4}
\end{equation}
where the values of $M_*$ for different accretion efficiencies can be calculated by using the numerical solutions of equations (30), (31) and (32). Here we also found that the initial masses of the presently evaporating PBHs vary inversely with radiation accretion efficiency. But matter accretion and vacuum energy accretion seem to have little effect on $\beta$ value. \\
Now using equation (37) with different values of $M_*$ corresponding to different values of radiation accretion efficiencies, we construct Table-5.
\begin{table}[h]
\centering
\begin{tabular}[c]{|c|c|c|}
\hline
\multicolumn{3}{|c|}{$t_{evap}=t_0$, $f_{mat}=0.35$ and $f_{vac}=0.6$}\\
\hline
$f_{rad}$ & $M_*$ & $\beta(M_*)<$ \\
\hline
0 & $2.3669\times10^{15}g$ & $3.832\times10^{-26}$\\
\hline
0.2 & $1.65685\times10^{15}g$ & $3.206\times10^{-26}$\\
\hline
0.4 & $0.94678\times10^{15}g$ & $2.424\times10^{-26}$\\
\hline
0.6 & $0.23669\times10^{15}g$ & $1.212\times10^{-26}$\\
\hline
\end{tabular}
\caption{Upper bounds on the initial mass fraction of PBHs those are evaporating  today for several accretion efficiencies $f$.}
\end{table}

If we neglect the presence of vacuum energy, we can obtain \cite{nsmprd1} that

\begin{equation}
\beta_0 (M_*)<\Big(\frac{M_*}{M_1}\Big)^{\frac{1}{2}} \Big(\frac{t_1}{t_0}\Big)^{\frac{2-n}{3}}\times 10^{-4}
\end{equation}
Comparing equation (37) and equation (38), we found
\begin{equation}
\beta(M_*)\sim 0.67\times \beta_0(M_*)
\end{equation}

It is clear from above equation that the constraint on the initial mass fraction of PBH obtained from the $\gamma$-ray background limit becomes a little stronger in the presence of vacuum energy but is nearly 1.8 times weaker compared with GTR result \cite{nj}.

\section{Conclusion}

In this paper, we study the evolution of primordial black holes within the context of Brans-Dicke theory by assuming present universe as vacuum dominated. We have considered the standard scenario of the universe passing through radiation domination and matter domination phases before being dominated by vacuum energy. In our analysis, we take the accretion of radiation, matter and vacuum energy only during respective dominant periods. We found that the rate of accretion of radiation is slower whereas the rate of accretion of matter is larger in BD theory in comparision of GTR. Though in contrast to GTR, here accretion of vacuum energy is possible throughout the PBH evolution during vacuum dominated era respecting the limit $ f_{vac}<0.61$, the rate of vacuum energy accretion is much slower than corresponding GTR result. Thus the PBHs evaporate at a faster rate in BD theory than Standard Cosmology if we consider the presence of vacuum energy in both cases. We also found that the constraint on the initial mass fraction of PBH obtained from the $\gamma$-ray background limit becomes stronger in the presence of vacuum energy as in Standard Cosmology though $1.8$ times weaker.

\end{document}